\documentclass{aa}  
\usepackage{graphicx}
\usepackage{epsfig}
\usepackage{txfonts}
\begin{document}
   \title{The chemical evolution of galaxies within the IGIMF theory: 
          the [$\alpha$/Fe] ratios and downsizing.}

   \author{S. Recchi
          \inst{1, 2}\thanks{simone.recchi@univie.ac.at}
          \and F. Calura\inst{3}\thanks{fcalura@oats.inaf.it}
          \and P. Kroupa\inst{4}\thanks{pavel@astro.uni-bonn.de}
          }

   \offprints{S. Recchi}

   \institute{Institute of Astronomy, Vienna University,
		T\"urkenschanzstrasse 17, A-1180, Vienna, Austria
        \and
              INAF - Osservatorio Astronomico di Trieste, 
                Via G.B. Tiepolo 11, 34143 Trieste, Italy
        \and
	      Astronomy Department, Trieste University, 
		Via G.B. Tiepolo 11, 34143 Trieste, Italy	
	\and
              Argelander Institute for Astronomy, Bonn University, 
              Auf dem H\"ugel 71, 53121 Bonn, Germany              
        }

   \date{Received; accepted}

 
  \abstract {The chemical evolution of galaxies is investigated within
  the framework of the star formation rate (SFR) dependent integrated
  galactic initial mass function (IGIMF).}  {We study how the global
  chemical evolution of a galaxy and in particular how [$\alpha$/Fe]
  abundance ratios are affected by the predicted steepening of the
  IGIMF with decreasing SFR.}{We use analytical and semi-analytical
  calculations to evaluate the mass-weighted and luminosity-weighted
  [$\alpha$/Fe] ratios in early-type galaxies of different
  masses.}{The models with the variable IGIMF produce a [$\alpha$/Fe]
  vs. velocity dispersion relation which has the same slope as the
  observations of massive galaxies, irrespective of the model
  parameters, provided that the star formation duration inversely
  correlates with the mass of the galaxy (downsizing).  These models
  also produce steeper [$\alpha$/Fe] vs. $\sigma$ relations in
  low-mass early-type galaxies and this trend is consistent with the
  observations.  Constant IMF models are able to reproduce the
  [$\alpha$/Fe] ratios in large elliptical galaxies as well, but they
  do not predict this change of slope for small galaxies.  In
  order to obtain the best fit between our results and the
  observations, the downsizing effect (i.e. the shorter duration of
  the star formation in larger galaxies) must be milder than
  previously thought.}{} 

   \maketitle
%

\section{Introduction}

It is nowadays widely accepted that most stars in galaxies form in
star clusters (Tutukov \cite{tutu78}; Lada \& Lada \cite{ll03}).  This
has been observed in a number of different galaxies; from the Milky
Way to the dwarf galaxies of the Local Group (Wyse et al. \cite{wyse};
Massey \cite{massey}; Piskunov et al. \cite{piskunov}).  Within each
star cluster, the initial mass function (IMF) can be well approximated
by the canonical two-part power-law form $\xi(m) \propto m^{-\alpha}$
(e.g. Pflamm-Altenburg, Weidner \& Kroupa \cite{pwk07}, hereafter
PWK07).  Massey \& Hunter (\cite{mh98}) have shown that for stellar
masses $m>$ a few M$_\odot$ a slope similar to the Salpeter
(\cite{salpeter}) index (i.e. $\alpha=2.35$) can approximate well the
IMF in clusters and OB associations for a wide range of metallicities,
whereas many studies have shown that the IMF flattens out below $m$
$\sim$ 1 M$_\odot$ (Kroupa, Tout \& Gilmore \cite{ktg93}; Chabrier
\cite{chab01}).

On the other hand, star clusters are also apparently distributed
according to a single-slope power law, $\xi_{\rm ecl} \propto M_{\rm
ecl}^{-\beta}$, where $M_{\rm ecl}$ is the stellar mass of the
embedded star cluster.  There is a general consensus that this slope
$\beta$ should be of the order of $\sim$ 2 (Zhang \& Fall \cite{zf99};
Lada \& Lada \cite{ll03}; Hunter et al. \cite{hunter}), although a
$\beta$ as high as 2.4 can also be realistic (Weidner, Kroupa \&
Larsen \cite{wkl04}).  According to this correlation, small embedded
clusters are more numerous in galaxies.  They provide therefore most
of the stars but not most of the massive ones, since they are
preferentially formed in massive clusters (Weidner \& Kroupa
\cite{wk06}).  As a consequence of this mass distribution of embedded
clusters, the integrated IMF in galaxies, the IGIMF, can be steeper
than the stellar IMF within each single star cluster (Kroupa \&
Weidner \cite{kw03}; Weidner \& Kroupa \cite{wk05}).

The Salpeter IMF slope has been used in a very wide range of modelling,
providing good fits with observations concerning the cosmic star
formation history (Calura, Matteucci \& Menci \cite{cmm04}), the X-ray
properties of elliptical galaxies (Pipino et al. \cite{pipino05}),
the chemical evolution of dwarf galaxies (Larsen, Sommer-Larsen \& Pagel
\cite{lslp}) and of the Milky Way (Pilyugin \& Edmunds \cite{pe06},
but see also Romano et al.  \cite{romano05}).  Broadly speaking, a
flatter than Salpeter IMF produces a larger fraction of massive stars.
The large production of oxygen (and of $\alpha$-elements in general)
leads to lower [Z/O] metallicity ratios.  A steep IMF slope would
instead be biased towards low- and intermediate-mass stars,
underproducing oxygen and therefore resulting in larger [N/O] and
[C/O] abundance ratios.  On the other hand, iron will also be
overproduced compared to $\alpha$-elements, since it comes mainly from
Type Ia SNe which originate from C-O deflagration of binary systems of
intermediate mass.  Therefore, galaxies characterized by a steep IMF
will tend to have [$\alpha$/Fe] ratios lower than models in which the
IMF is flat.

The scenario of a variable integrated galactic initial mass function
(IGIMF) has been applied in models of chemical evolution (K\"oppen,
Weidner \& Kroupa \cite{kwk07}), producing an excellent agreement with
the mass-metallicity relation found by Tremonti et al. (\cite{t04}).
However, these authors consider only the effect of the IGIMF on the
global metallicity and the evolution of abundance ratios has not yet
been explored in the literature.  In a series of papers we plan to
study the impact of the IGIMF on the abundance ratios in different
classes of galaxies, using different methodologies.  In this paper we
study, by means of simple analytical and semi-analytical models, the
evolution of [$\alpha$/Fe] ratios in galaxies, in particular in
early-type ones.  It is, in fact, now well established that the
[$\alpha$/Fe] ratios in the cores of elliptical galaxies increase with
galactic mass (Weiss, Peletier \& Matteucci \cite{wpm95}; Kuntschner
et al. \cite{kunt01}) and this poses serious problems to the current
paradigm of hierarchical build-up of galaxies (see e.g. Thomas et
al. \cite{thom05}, hereafter THOM05; Nagashima et al. \cite{naga05};
Pipino, Silk \& Matteucci \cite{psm09}; Calura \& Menci, in
preparation).  In fact, in the classical hierarchical models the most
massive ellipticals take a longer time to assemble and therefore form
stars for a longer time than less massive galaxies, thus producing a a
trend of [$\alpha$/Fe] vs. mass which is opposite of what is observed
(see Thomas, Maraston \& Bender \cite{thom02}; Matteucci
\cite{matt07}).

We will show in this paper that the trend of increasing [$\alpha$/Fe]
vs. galaxy mass is naturally accounted for in models of elliptical
galaxies in which the IGIMF is implemented.  The second paper of this
series will be devoted to the study of the chemical evolution of the
Solar Neighborhood and of the local dwarf galaxies and in this case we
will make use of detailed chemical evolution models.  Another paper of
this series will study the evolution of galaxies by means of
chemodynamical models, in order to analyze how the IGIMF changes the
feedback of the ongoing star formation in galaxies and how this
affects the chemical evolution.

The plan of the present paper is as follows. In Sect. 2 we summarize
the IGIMF theory and the formulations we adopt.  In Sect. 3 we
describe how we calculate the Type Ia and Type II SN rates in galaxies
in which the SFR is given.  Once we know the Type Ia and Type II SN
rates, it is possible to calculate the [$\alpha$/Fe] ratios.  This has
been done in Sect. 4 for ellipticals and early-type galaxies in
general.  A discussion and the main conclusions are presented in
Sect. 5.


\section{The determination of the integrated galactic initial mass function}
\label{sec:igimf}

The determination of the IGIMF has been described previously (Kroupa
\& Weidner \cite{kw03}; Weidner \& Kroupa \cite{wk05}; PWK07).  
The IGIMF theory is based on the assumption that all the stars in
a galaxy form in star clusters.  Surveys of star-formation in the
local Milky Way disk have shown that 70 to 90 \% of all stars appear
to form in embedded clusters (Lada \& Lada \cite{ll03}; Evans et
al. \cite{evans08}). The remaining 10-30 \% of the apparently
distributed population may stem from a large number of short-lived
small clusters that evolve rapidly by dissolving through energy
equipartition and residual gas expulsion. It is therefore reasonable
to assume that star formation occurs in embedded clusters with masses
ranging from a few M$_\odot$ upwards.  The IGIMF, integrated over the
whole population of embedded clusters forming in a galaxy, becomes

\begin{equation}
\xi_{\rm IGIMF}(m;{\rm SFR} (t)) = 
\int_{M_{\rm ecl, min}}^{M_{\rm ecl, max} ({\rm SFR} (t))} 
\hspace{-0.6cm}\xi (m \leq m_{\rm max}) \xi_{\rm ecl} (M_{\rm ecl}) 
d M_{\rm ecl},
\end{equation}
\noindent
where $M_{\rm ecl, min}$ and $M_{\rm ecl, max} ({\rm SFR} (t))$ are
the minimum and maximum possible masses of the embedded clusters in a
population of clusters and $m_{\rm max} = m_{\rm max} (M_{\rm ecl})$
(eqs. \ref{eq:normmecl} and \ref{eq:normxi}).  For $M_{\rm ecl, min}$
we take 5 M$_\odot$ (the mass of a Taurus-Auriga aggregate, which is
arguably the smallest star-forming "cluster" known), whereas the upper
mass of the embedded cluster population depends on the SFR.  The
correlation between $M_{\rm ecl, max}$ and SFR has been determined
observationally (Larsen \& Richtler \cite{lr00}; Weidner et
al. \cite{wkl04}) and can be expressed with the correlation

\begin{equation}
\log M_{\rm ecl, max} = \log k_{\rm ML} + 0.75 \log \psi + 6.77,
\label{eq:mecl}
\end{equation}
where $\psi$ is the SFR in M$_\odot$ yr$^{-1}$ and $k_{\rm ML}$ is the
mass-to-light ratio, typically 0.0114 for young stellar populations
(Smith \& Gallagher \cite{sg01}).  This empirical finding can be
understood to result from the sampling of clusters from the embedded
cluster mass function given the amount of gas mass being turned into
stars per unit time (Weidner et al. \cite{wkl04}).

The stellar IMF (i.e. the IMF within each embedded cluster) has the
canonical form $\xi(m) = k m^{-\alpha}$, with $\alpha = 1.3$ for 0.08
M$_\odot \le$ $m$ $<$ 0.5 M$_\odot$ and $\alpha = 2.35$ (i.e. the
Salpeter slope) for 0.5 M$_\odot \le$ $m$ $< m_{\rm max}$, where $m_{\rm
max}$ depends on the mass of the embedded cluster.  In order to
determine $m_{\rm max}$ and the proportionality constant $k$ we have
to solve the following two equations (Kroupa \& Weidner \cite{kw03}):

\begin{equation}
M_{\rm ecl} = \int_{m_{\rm low}}^{m_{\rm max}} m \xi (m) dm,
\label{eq:normmecl}
\end{equation}

\begin{equation}
\int_{m_{\rm max}}^{m_{\rm max *}} \xi (m) dm = 1,
\label{eq:normxi}
\end{equation}
\noindent
where $m_{\rm low}$ is the smallest considered stellar mass (0.08
M$_\odot$ in our case) and $m_{\rm max *}$ is the upper physical
stellar mass and its value is assumed to be 150 M$_\odot$ (Weidner \&
Kroupa \cite{wk04}).  Eq. \ref{eq:normxi} indicates that, by
definition of $m_{\rm max}$, there is only one and exactly one star in
the embedded cluster with mass $M_{\rm ecl}$ whose mass is larger than
or equal to $m_{\rm max}$.

\begin{figure}
\centering
\epsfig{file=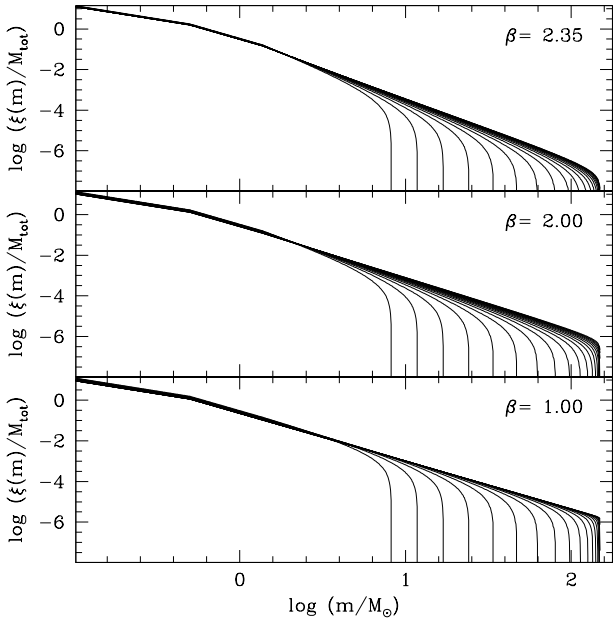, height=9cm,width=9cm}
\caption[]{\label{igimf_ext} IGIMFs for different distributions of 
embedded clusters (e.g. different values of $\beta$).  Upper panel: 
model BETA235; central panel: model BETA200; lower panel: model 
BETA100.  For each panel we have considered 20 possible values of 
SFR, ranging from 10$^{-4}$ M$_\odot$ yr$^{-1}$ (lowermost curves) 
to 100 M$_\odot$ yr$^{-1}$ (uppermost curves), equally spaced in
logarithm.}
\end{figure}

\begin{figure}
\centering
\epsfig{file=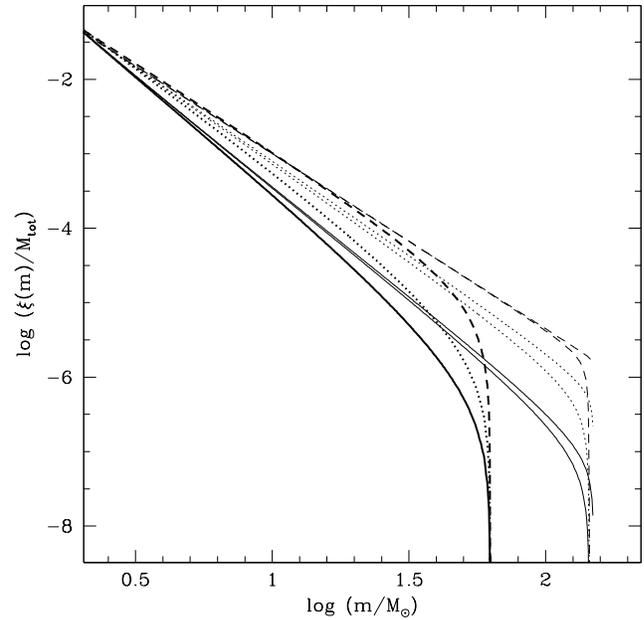, height=9cm,width=9cm}
\caption[]{\label{igimf_comp} IGIMFs for different SFRs: 
$\simeq$ 10$^{-2}$ M$_\odot$ yr$^{-1}$ (heavy lines); $\simeq$ 1
M$_\odot$ yr$^{-1}$ (middle lines); $\simeq$ 10$^{2}$ M$_\odot$
yr$^{-1}$ (light lines).  We have considered all the possible values
of $\beta$: model BETA100 (dashed lines), BETA200 (dotted lines),
BETA235 (solid lines). }
\end{figure}

The last ingredient we need is the distribution function of embedded
clusters, $\xi_{\rm ecl} (M_{\rm ecl})$, which, as we have mentioned
in the Introduction, we can assume proportional to $M_{\rm
ecl}^{-\beta}$.  In this work we have assumed 3 possible values of
$\beta$: 1.00 (model BETA100), 2.00 (model BETA200) and 2.35 (model
BETA235).  In Fig. \ref{igimf_ext} we have plotted the resulting
IGIMFs for different values of SFR.  In particular, we have tested 20
SFRs, ranging from 10$^{-4}$ to 100 M$_\odot$ yr$^{-1}$, equally
spaced in logarithm.  To appreciate better the differences between
various models, we have plotted in Fig. \ref{igimf_comp} IGIMFs for 3
different values of the SFR: SFR $\simeq$ 10$^{-2}$ M$_\odot$
yr$^{-1}$ (heavy lines), SFR $\simeq$ 1 M$_\odot$ yr$^{-1}$ (middle
lines), SFR $\simeq$ 10$^{2}$ M$_\odot$ yr$^{-1}$ (light lines).  We
have considered all the possible values of $\beta$: model BETA100
(dashed lines), BETA200 (dotted lines), BETA235 (solid lines).  For
clarity, we have plotted the IGIMFs only for masses larger than $\sim$
2 M$_\odot$, since in the range of low mass stars the IGIMFs do not
vary.  As expected, the model with the steepest distribution of
embedded clusters (model BETA235) produces also the steepest IGIMFs.
This is due to the fact that model BETA235 is biased towards embedded
clusters of low mass, therefore the probability of finding high mass
stars in this cluster population is lower.

\begin{figure}
\centering
\epsfig{file=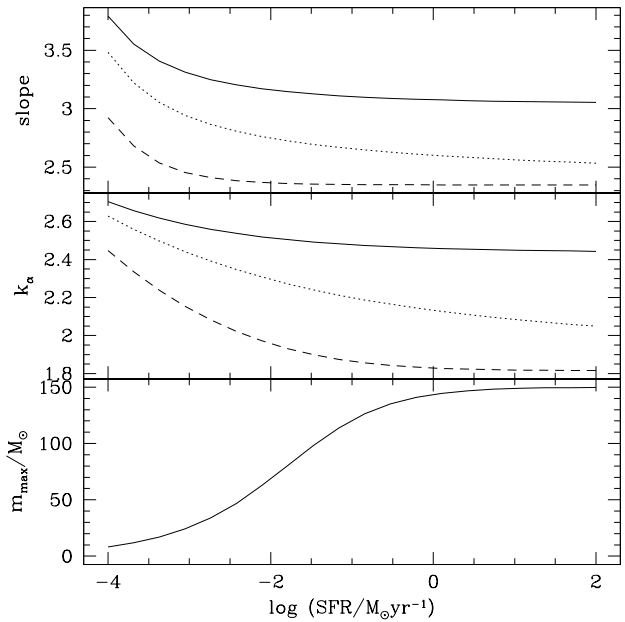, height=9cm,width=9cm}
\caption[]{\label{slope} Lower panel: $m_{\rm max}$ (in M$_\odot$) as 
a function of SFR (in M$_\odot$ yr$^{-1}$).  Middle panel:
$k_{\alpha}$ (number of stars per unit mass in one stellar
generation).  Upper panel: IGIMF slopes in the range 3 - 16 M$_\odot$.
Notations as in Fig. \ref{igimf_comp}. }
\end{figure}

We can also notice from Fig. \ref{igimf_comp} that the differences
between IGIMFs with SFR $\simeq$ 1 M$_\odot$ yr$^{-1}$ (middle lines)
and SFR $\simeq$ 10$^{2}$ M$_\odot$ yr$^{-1}$ (light lines) are not
very pronounced.  This is due to the fact that for both these SFRs, 
the maximum possible mass of the embedded cluster is very high (see eq. 
\ref{eq:mecl}), therefore in both cases the upper possible stellar mass of 
the whole galaxy is very close to the theoretical limit of 150
M$_\odot$.  This can be seen in Fig. \ref{slope} (lower panel) in
which we plot the variation of $m_{\rm max}$ as a function of SFR as
deduced from eqs. \ref{eq:normxi} and \ref{eq:normmecl}.  This
correlation is valid for all the possible values of $\beta$ because it
is determined by $\xi (m)$ and not by $\xi_{\rm ecl} (M_{\rm ecl})$.
As we can see from Figs. \ref{igimf_ext} and \ref{igimf_comp}, the
IGIMFs are characterized by a nearly uniform decline, which follows
approximately a power law, and a sharp cutoff when $m$ gets close to
$m_{\rm max}$.  In Fig. \ref{slope} (upper panel) we plot therefore
also the slope that better approximates the IGIMF in the range 3 - 16
M$_\odot$.  This is the range of masses where most of the progenitors
of SNeII and SNeIa originate (see Sect. 3).  Of course, the steeper
the distribution of embedded clusters is, the steeper the
corresponding IGIMFs are.  Fig. \ref{slope} shows also what we have
noticed before, namely that the various IGIMFs saturate for SFR $>$ 1
M$_\odot$ yr$^{-1}$.  Finally, in Fig. \ref{slope} (middle panel)
$k_{\alpha}$ is shown as a function of the SFR for the various models.
$k_{\alpha}$ is the number of stars per unit mass in one stellar
generation (see e.g. Greggio \cite{greggio05}) and its value is given
by 

\begin{equation}
k_{\alpha} = {{\int_{m_{\rm low}}^{m_{\rm max}} \xi_{\rm IGIMF} (m) dm}
\over
{\int_{m_{\rm low}}^{m_{\rm max}} m \xi_{\rm IGIMF} (m) dm}}.  
\end{equation}
\noindent
This parameter is useful to calculate the SNII rates (see Sect. 3).

\section{The determination of Type Ia and Type II SN rates}

\subsection{Type II SN rates}
\label{subs:SNII}

Stars in the range $m_{\rm up} < m < m_{\rm max}$ (where $m_{\rm
up}$ is the mass limit for the formation of a degenerate C--O core)
are generally supposed to end their lives as {\it core-collapse} SNe.
These SNe divide into SNeII, SNeIb and SNeIc according to their
spectra.  For our purposes, this distinction is not useful and we will
suppose that all the core-collapse supernovae are indeed SNeII.  These
SNe produce the bulk of $\alpha$-elements and some iron (one third
approximately).  The standard value of $m_{\rm up}$ is 8 M$_\odot$ but
stellar models with overshooting predict lower values (e.g. Marigo
\cite{marigo01}).  However, stars more massive than  $m_{\rm up}$ can 
still develop a degenerate O--Ne core and end their lives as
electron-capture SNe (Siess \cite{siess07}).  We will assume for 
simplicity that all the stars with masses larger than 8~M$_\odot$ end 
their lives as SNeII, therefore the SNII rate is simply given by the 
rate at which massive stars die, namely:

\begin{equation}
R_{SNII} (t) = \int_8^{m_{\rm max}} \psi (t - \tau_m) \xi_{\rm IGIMF} [m, \psi 
(t - \tau_m)] dm,
\label{eq:snii}
\end{equation}
\noindent
where $\psi$ is the SFR and $\tau_m$ is the lifetime of a star of mass
$m$.  The lifetime function is adapted from the work of Padovani \&
Matteucci (\cite{pm93}), 

\begin{equation}
\tau_m = \cases{1.2 m^{-1.85}+0.003\;{\rm Gyr} &if $m\geq 6.6$ M$_{\odot}$\cr
     10^{f(m)}
\;{\rm Gyr} &if $m<6.6$ M$_{\odot}$,\cr}
\end{equation}
\noindent
where 

\begin{equation}
f(m)={{\bigl\lbrack 0.334-\sqrt{1.79-0.2232\times(7.764-\log(m))}
\bigr\rbrack\over 0.1116}}.
\end{equation}
\noindent
In eq. \ref{eq:snii} the IGIMF is calculated by considering the SFR at
the time $t - \tau_m$, therefore it depends both on time and on mass.

\begin{figure}
\centering
\epsfig{file=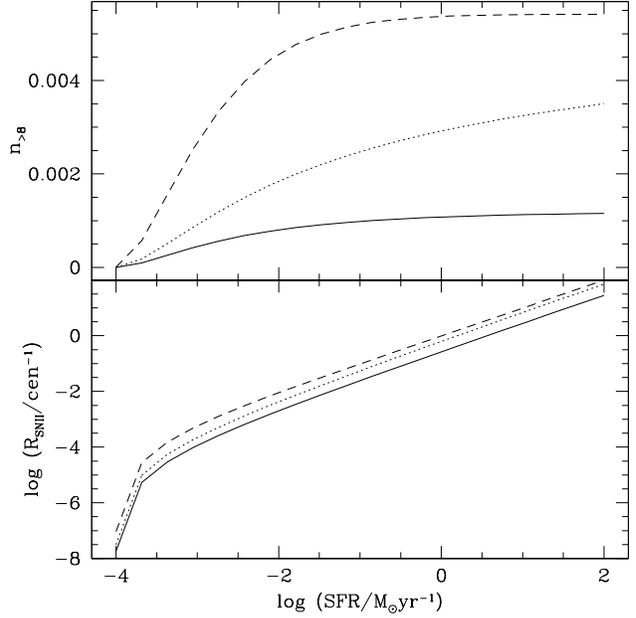, height=9cm,width=9cm}
\caption[]{\label{sn2_cen} Lower panel: $R_{SNII}$ (in cen$^{-1}$) as a 
function of SFR (in M$_\odot$ yr$^{-1}$).  Upper panel: $n_{>8}$
(number fraction of stars more massive than 8 M$_\odot$) as a function
of SFR.  Notation as in Fig. \ref{igimf_comp}.}
\end{figure}

It is instructive to analyze models in which SFR is constant during
the whole evolution of the galaxy.  In this way, eq. \ref{eq:snii}
simplifies into

\begin{equation}
R_{SNII} = \psi \int_8^{m_{\rm max}} \xi_{\rm IGIMF} (m, \psi) dm = 
\psi k_\alpha n_{> 8},
\label{eq:r2}
\end{equation}
\noindent
where, as we have seen in Sect. 2, $k_{\alpha}$ is the number of stars
per unit mass in one stellar generation (therefore $\psi k_\alpha$ is
the number of stars formed per unit time) and $n_{> 8}$ is the number
fraction of stars with masses larger than 8 M$_\odot$.  In
Fig. \ref{sn2_cen} (lower panel) we plot $R_{SNII}$ (in cen$^{-1}$) as
a function of SFR for the 3 adopted values of $\beta$.  It is worth
pointing out that, even if SFR is constant, $R_{SNII}$ starts
increasing only after the star with mass $m_{\rm max}$ ends its life
and reaches a constant value only after the lifetime of a 8 M$_\odot$
star.  This lifetime ($\sim$ 28 Myr with the adopted lifetime
function) is however negligible compared to the Hubble time, therefore
it is reasonable to consider $R_{SNII}$ constant with time.  For SFRs
larger than $\sim$ 10$^{-2}$ M$_\odot$ yr$^{-1}$ $R_{SNII}$ increases
almost monotonically with SFR, whereas it drops dramatically for SFR
$<$ 10$^{-2}$ M$_\odot$ yr$^{-1}$.  This is mostly due to the drop of
$n_{>8}$ at low SFRs (Fig. \ref{sn2_cen}, upper panel) which in turn
depends on the fact that the upper mass $m_{\rm max}$ for these values
of SFR starts reducing significantly and it gets very close to 8
M$_\odot$ for a SFR of 10$^{-4}$ M$_\odot$ yr$^{-1}$
(Fig. \ref{slope}), therefore only a very narrow interval of stellar
masses gives rise to SNII explosions.  From Fig. \ref{slope} we can
see instead that the variation of $k_{\alpha}$ with SFR is not very
significant, therefore $k_{\alpha}$ affects only mildly the Type II SN
rates.

It is nowadays getting popular to consider SN rates normalized to the
stellar mass of the considered galaxy.  The usually chosen unit of
measure is the SNuM (1 SNuM = 1 SN cen$^{-1}$ 10$^{-10}$ M$_*$$^{-1}$,
where M$_*$ is the current stellar mass of the galaxy).  In this case,
models in which the SFR is constant cannot attain a constant
Type~II~SN rate in SNuM since the stellar mass of the galaxy increases
with time.  We therefore calculated $R_{SNII}$ in SNuM as a function
of time for the various models.  The stellar mass of the galaxy at
each time $t$ is given by $\int_0^t \psi \cdot f_{< m (t)} dt$, where
$f_{< m (t)}$ is the mass fraction of stars, born until the time $t$
that have not yet died.

\begin{figure}
\centering
\epsfig{file=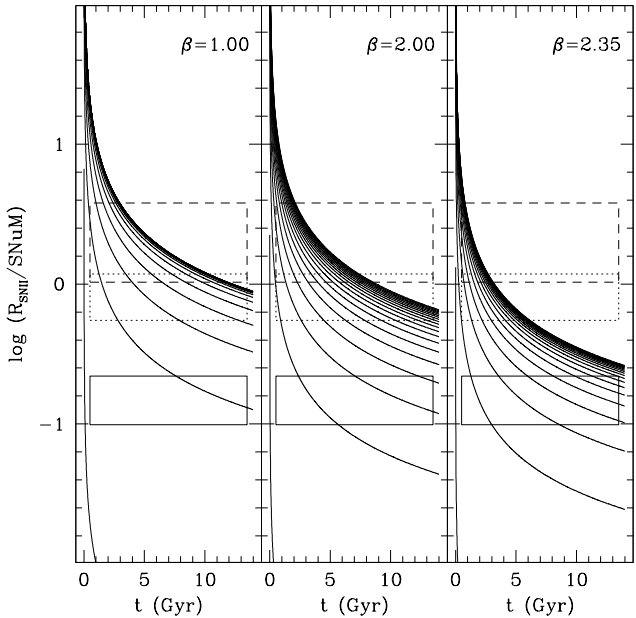, height=9cm,width=9cm}
\caption[]{\label{sn2_snum} $R_{SNII}$ (in SNuM) as a function 
of time (in Gyr) for models with constant SFR and different values of
$\beta$: model BETA100 (left panel), BETA200 (central panel), BETA235
(rigth panel).  In each panel, the lowermost curve has SFR = 10$^{-4}$
M$_\odot$ yr$^{-1}$, the uppermost has SFR = 10$^{2}$ M$_\odot$
yr$^{-1}$ and the other SFRs are equally spaced in logarithm.
Observations of the core-collapse SN rates in S0a/b galaxies (solid
boxes), in Sbc/d galaxies (dotted boxes) and in irregular ones
(short-dashed boxes) are also shown.  Data are taken from Mannucci et
al. (\cite{mann05})}
\end{figure}

Fig.  \ref{sn2_snum} shows the evolution with time of the Type II SN
rate for different models and different SFRs, assuming a constant SFR
for 14 Gyr.  These results are compared with the average SNeII rates
(in SNuM), observationally derived by Mannucci et al. (\cite{mann05})
in S0a/b galaxies (solid boxes), Sbc/d galaxies (dotted boxes) and
irregular ones (dashed boxes).  We can notice that, for the models
BETA100 and BETA200 only the mildest SFRs can reproduce the final SNII
rates in S0a/b galaxies, whereas model BETA235 can fit the final SNII
rate of S0a/b galaxies for a wide range of SFRs.  On the other hand,
all the models predict final SNII rates significantly below the
observations of irregular galaxies and the final values for model
BETA235 fail also to fit the observed rates in Sbc/d galaxies.  It is
important to note, however, that the stellar mass in galaxies is
usually calculated assuming some (constant) IMF.  Under the assumption
that the IMF changes with the SFR, the determinations of the stellar
masses must be revisited.  PWK07 showed that the IGIMF effect
(i.e. the suppression of the number of massive stars with respect of
low-mass stars) can be very significant in dwarf galaxies, whereas in
large galaxies it tends to be very small.  Moreover, a constant SFR
for 14 Gyr is not a reasonable description of the star formation
history of irregular (and Sbc/d) galaxies which often experience an
increase of the SFR in the last Gyrs of their evolution (see
e.g. Calura \& Matteucci \cite{cm06}).  For this reason, the
calculated SNII rates of late type galaxies tend to fit the
observations at smaller ages.

\subsection{Type Ia SN rates}
\label{subs:SNIa}

In order to calculate the SNIa rates, we assume the so-called Single
Degenerate Scenario of SNIa formation.  It is commonly assumed that a
SNIa explodes when a C-O white dwarf in a binary system reaches the
Chandrasekhar mass after mass accretion from a companion star.
According to the Single Degenerate channel of SNIa explosion, the
accretion of matter occurs via mass transfer from a non-degenerate
companion (a red giant or a main sequence star) filling its Roche lobe
(Whelan \& Iben \cite{wi73}).  In this way, the SNIa rate depends on
the number distribution of C-O white dwarfs, but also on the mass
ratio between primary and secondary stars in a binary system.  The
SNIa rate in the framework of the Single Degenerate Scenario has been
analytically calculated by a number of authors assuming a universal
IMF (see Valiante \cite{vali09} and references therein).  Here we
follow the formulation of Greggio \& Renzini (\cite{gr83}) and
Matteucci \& Recchi (\cite{mr01}) but we modify it to take into
account that, in the framework of the IGIMF, the IMF changes according
to the SFR.  The SNIa rate in this case turns out to be:

\begin{equation}
R_{\rm Ia} (t)=A \hspace{-0.1cm}\int_{m_{\rm B, inf}}^{m_{\rm B, sup}}
\hspace{-0.25cm}\int_{\mu_{\rm min}}^{0.5}\hspace{-0.25cm}f(\mu)
\psi(t-\tau_{m_2})\xi_{\rm IGIMF} [m_{\rm B}, \psi 
(t - \tau_{m_2})]d\mu\, dm_{\rm B},
\label{eq:r1a}
\end{equation}
\noindent
where $A$ is a normalization constant (assumed to be 0.09 in the
following).  Although theoretical arguments demonstrate that $A$
should be small (e.g. Maoz \cite{maoz08}) its value is usually
calibrated with the Milky Way.  Unfortunately, our analytical approach
does not allow us to simulate the Milky Way within the IGIMF theory,
therefore we take 0.09 as a reference value and postpone a more
careful discussion about it to the follow-up numerical paper (but see
also Sec. \ref{sec:results} for a study of the variation of A for
early-type galaxies).  $m_{\rm B}$ is the total mass of the binary
system, $m_2$ is the mass of the secondary star, $\mu = m_2 / m_{\rm
B}$ and $f(\mu)$ is the distribution function of mass ratios (see
below).  It is commonly assumed that the maximum stellar mass able to
produce a degenerate C-O white dwarf is a 8~M$_\odot$ star, therefore
the maximum possible binary mass is 16 M$_\odot$.  The minimum
possible binary mass is assumed to be 3 M$_\odot$ in order to ensure
that the smallest possible white dwarf can accrete enough mass from
the secondary star to reach the Chandrasekhar mass.  With these
assumptions, the limits of integration in eq. (\ref{eq:r1a}) are:

\begin{equation}
m_{\rm B, inf} = max [2 m_2 (t), 3 {\rm M}_\odot]
\end{equation}

\begin{equation}
m_{\rm B, sup} = 8 {\rm M}_\odot + m_2 (t),
\end{equation}

\begin{equation}
\mu_{\rm min} = max \biggl[{m_2 (t) \over m_B}, 
{{m_B - 8 {\rm M}_\odot}\over m_B}\biggr].
\end{equation}
\noindent
The distribution function of mass ratios is generally described as a
power law ($f(\mu) \propto \mu^{\gamma}$), but the value of $\gamma$
is still much debated in the literature (Duquennoy \& Mayor
\cite{dm91}; Shatsky \& Tokovinin \cite{st02}; Kouwenhoven et
al. \cite{kouw05}) and therefore we will take it as a free
parameter. \footnote{Note that usually observative papers adopt mass
ratios $q = m_2/m_1$ instead of $\mu$ (Abt \& Levy \cite{al76}).  The
mass ratio distribution function $f (q)$ can be obtained from $f
(\mu)$ by means of a simple change of variable.}

\begin{figure}
\centering
\epsfig{file=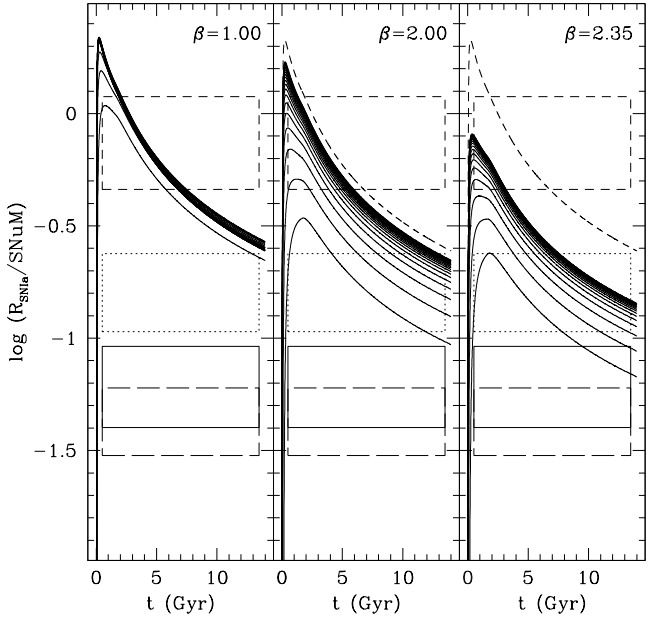, height=9cm,width=9cm}
\caption[]{\label{sn1a_snum} $R_{SNIa}$ (in SNuM) as a function 
of time (in Gyr) for models with constant SFR.  Notations and symbols
as in Fig. \ref{sn2_snum}, with the addition of the observed SNIa rate
in E/S0 galaxies (long-dashed boxes) and the SNIa rates predicted
assuming a constant IMF (dashed lines).  Data are taken from Mannucci
et al. (\cite{mann05}).}
\end{figure}

\begin{figure}
\centering
\epsfig{file=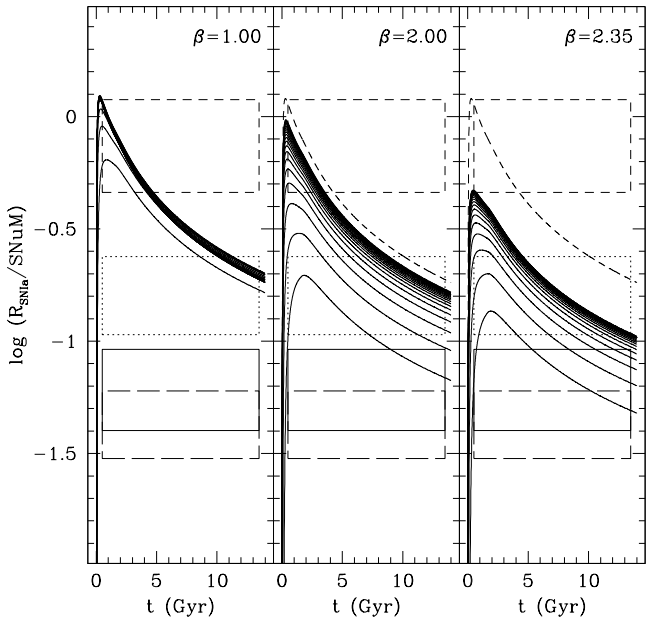, height=9cm,width=9cm}
\caption[]{\label{sn1a_snum_0.3} As in Fig. \ref{sn1a_snum} but for 
$\gamma = 0.3$.}
\end{figure}

Fig.  \ref{sn1a_snum} shows the evolution with time of the Type Ia SN
rate for different models and different SFRs, analogously to
Fig. \ref{sn2_snum} for SNeII rates.  Also shown (dashed lines)
for comparison are SNIa rates obtained for a model with fixed
(i.e. not SFR-dependent) IMF.  We assume the canonical stellar IMF
(i.e. the IMF within each embedded cluster) which, as mentioned in
Sect. \ref{sec:igimf}, has the form $\xi(m) = k m^{-\alpha}$, with
$\alpha = 1.3$ for 0.08 M$_\odot <$ $m$ $<$ 0.5 M$_\odot$ and $\alpha
= 2.35$ above 0.5 M$_\odot$.  As we can see, at large SFRs model
BETA100 produces rates almost indistinguishable from the ones obtained
with the fixed canonical IMF (see also Kroupa \& Weidner \cite{kw03}).
In this figure $\gamma$ is assumed to be 2 (Tutukov \& Yungelson
\cite{ty80}).  This large value of $\gamma$ favors the occurrence of
SNeIa in binary systems with similar masses.  Such a steep mass ratio
distribution that favors equal-mass binaries may result from dynamical
evolution of stellar populations in long-lived star clusters (Shara \&
Hurley \cite{sh02}).  We can notice again that only model BETA235 at
very low SFRs seems able to reproduce the SNIa rates in S0a/b and E/S0
galaxies.  However, we point out that the comparison with the observed
SNIa rates in elliptical galaxies is meaningless because they have
stopped forming stars several Gyr ago, therefore they have evolved
passively since then.  For them we cannot therefore assume a constant
SFR for 14 Gyr (see Sect. \ref{sec:results}).  On the other hand,
model BETA235 produces SNIa rates that only match the observed rates
in dwarf irregular galaxies at their peak.  Therefore, assuming
$\gamma = 2$, the best value for $\beta$ seems to be 2 (but see the
comment in Sect. \ref{subs:SNII} about the possible inconsistency of
the published determination of stellar masses, at least for irregular
galaxies).  To show the dependence of the results on $\gamma$ we show
in Fig. \ref{sn1a_snum_0.3} the SNeIa rates obtained assuming $\gamma
= 0.3$.  This flatter distribution function implies that a larger
fraction of binary systems with small mass ratios end up as SNeIa.  We
can notice from this figure that the observed SNIa rates in spiral
galaxies are reproduced by a larger range of SFRs, whereas the
disagreement with the observed rates in irregular galaxies worsens.

\section{A test of the IGIMF: [$\alpha$/Fe] ratios in early-type 
galaxies}
\label{sec:results}

The study of the average stellar [$\alpha$/Fe] ratio in galaxies
represents an important constraint for our models, since this quantity
depends both on the adopted galactic star formation history and on the
stellar IMF (Matteucci \cite{mat01}).  In local ellipticals, the
observed correlation between the central velocity dispersion $\sigma$,
which reflects the total stellar mass, and the stellar [$\alpha$/Fe]
is interpreted as due to the shorter star formation timescales in the
most massive galaxies (Pipino \& Mattuecci \cite{pm04}; THOM05) which
in turn implies also that the most massive galaxies experience the
most intense episodes of star formation. For this reason, the average
stellar [$\alpha$/Fe] vs $\sigma$ relation represents a valuable test
for the IGIMF, since the IGIMF is a function of the galactic star
formation rate.  The issue of a variable IMF among elliptical galaxies
to explain the [$\alpha$/Fe] vs. $\sigma$ relation has already been
explored with success by Matteucci (\cite{matt94}) but assuming ad hoc
variations of the IMF slope.  In this section we test, using
well-established and observationally constrained star formation
histories of early-type galaxies of various masses, if the physically
motivated IGIMF can equally well reproduce this correlation.

\begin{figure}
\centering
\epsfig{file=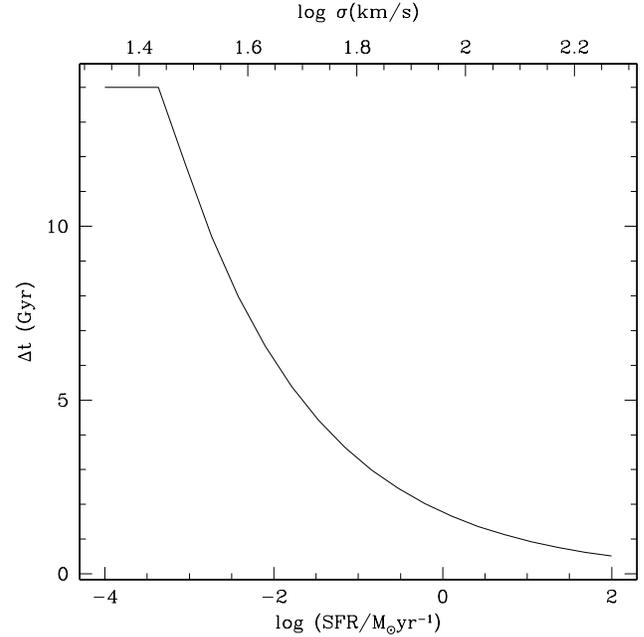, height=9cm,width=9cm}
\caption[]{\label{deltat} Duration $\Delta t$ of the star formation 
(assumed constant) as a function of SFR (lower scale) and $\sigma$
(upper scale) assuming the $\Delta t$-luminous mass relation of THOM05
(see text).}
\end{figure}

To simplify the calculations, the SFR is assumed to be constant over a
period of time $\Delta t$.  We have numerically tested that this crude
approximation about the star formation history does not affect
significantly the results.  The value of $\Delta t$ as a function
of galaxy luminous mass is adopted from the work of THOM05, who, on
the basis of the observational relation between [$\alpha$/Fe] and
$\sigma$, showed the existence of a {\it downsizing} pattern for
elliptical galaxies, according to which the smaller ellipticals form
over longer timescales (see also Matteucci \cite{matt94}; Cowie et
al. \cite{cowie96}; Kodama et al. \cite{koda04}).  Since the
present-day stellar mass is given in this case by $\psi \cdot
\int_o^{\Delta t}$ $f_{\rm low} (t) dt$ (where $f_{\rm low} (t)$ is
the fraction of long-living stars, namely the stars, born at the time
$t$, that live until the present day), it is possible to derive a
relation between the SFR and the duration of the star formation
activity $\Delta t$, which we show in Fig. \ref{deltat}.  This
relation saturates at 14 Gyr since this is assumed to be the age of
the Universe.  A similar relation can be recovered from the work of
Pipino \& Matteucci (\cite{pm04}) assuming that the star formation
occurs only until the onset of the galactic wind, however the two
SFR-$\Delta t$ relations do not significantly differ.

\begin{figure}
\centering
\epsfig{file=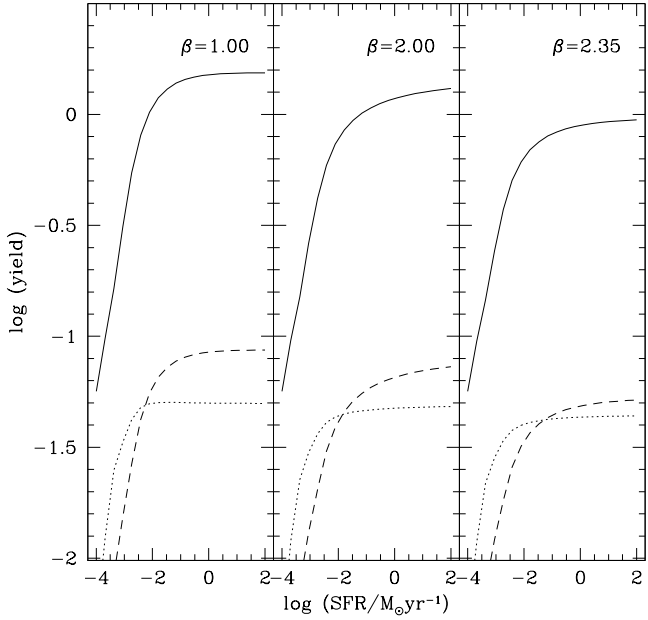, height=9cm,width=9cm}
\caption[]{\label{yields} IGIMF- averaged SNII yields of oxygen (solid
lines), iron (dotted lines) and magnesium (dashed lines) as a function
of SFR (in M$_\odot$ yr$^{-1}$) for different values of $\beta$: model
BETA100 (left panel), BETA200 (central panel), BETA235 (right panel).}
\end{figure}

For each galaxy (characterized by a specific SFR over a period $\Delta
t$) we calculate the average yield from SNeII of a chemical element $i$,

\begin{equation}
\overline{y^{\rm II}_{\rm i}} = 
{{\int_8^{m_{\rm max}} y_{\rm i} (m) \xi_{\rm IGIMF} (m, \psi) dm}
\over {\int_8^{m_{\rm max}} \xi_{\rm IGIMF} (m, \psi) dm}},
\end{equation}
\noindent
where $y_{\rm i} (m)$ is the yield of chemical element $i$ produced by
a single star of mass $m$.  The nucleosynthetic prescriptions are
taken from Woosley \& Weaver (\cite{ww95}).  We have however halved
the iron yields, in accordance with Timmes, Woosley \& Weaver
(\cite{tww95}) and Chiappini, Matteucci \& Gratton (\cite{cmg97}),
because it is known that only in this way it is possible to reproduce
the [$\alpha$/Fe] in Galactic stars.  Unfortunately, Woosley \& Weaver
(\cite{ww95}) calculated yields only for stellar masses up to 40
M$_\odot$.  We assume that the yields of stars more massive than 40
M$_\odot$ are equal to the 40 M$_\odot$ star yields.  Given the very
limited amount of stars in the range 40 M$_\odot$ $<$ $m$ $<$ $m_{\rm
max}$ the results are not sensitive to this assumption.  In
Fig. \ref{yields} we show the IGIMF-averaged SNII yields of oxygen
(solid lines), iron (dotted lines) and magnesium (dashed lines) as a
function of the SFR for different values of $\beta$.  For what
concerns SNeIa, we assume the yields reported by Gibson, Loewenstein
\& Mushotzky (\cite{glm97}), based on the work of Thielemann, Nomoto
\& Hashimoto (\cite{tnh93}).

Once we know the SNIa yields and the IGIMF-averaged SNII yields for
each galaxy, we can calculate the mass fraction ${\alpha \over Fe}
(t)$ (where $\alpha$ is O or Mg) produced until the time $t$ by using
the formula:

\begin{equation}
{\alpha \over Fe} (t) = {{\int_0^t \bigl(R_{\rm Ia} (t) y_\alpha^{\rm
Ia} + R_{\rm SNII} \overline{y^{\rm II}_\alpha}\bigr) dt} \over
{\int_0^t \bigl(R_{\rm Ia} (t) y_{Fe}^{\rm Ia} + R_{\rm SNII}
\overline{y^{\rm II}_{Fe}}\bigr) dt} },
\label{eq:alphafe_gas}
\end{equation}
\noindent
where $R_{\rm Ia}$ and $R_{\rm SNII}$ are the SNIa and SNII rates,
given by eqs. \ref{eq:r1a} and \ref{eq:r2}, respectively, and $y^{\rm
Ia}$ are the SNIa yields.

At this point, we can compute the theoretical average stellar
abundances by means of:

\begin{equation}
[\alpha/Fe] = \log_{10} \frac{\psi \cdot \int_{0}^{\Delta t} 
{\alpha \over Fe} (t) 
\cdot f_{\rm low} (t) dt}{M_{tot}} - \log_{10}{\alpha_\odot
\over Fe_\odot}.
\end{equation}
\noindent
where $M_{tot}$ is the total present-day stellar mass and
$\alpha_\odot$ and $Fe_\odot$ are the solar abundances of
$\alpha$-elements and Fe, respectively, taken from Anders \& Grevesse
(\cite{ag89}).  This means that our theoretical abundance ratios
represent values mass-averaged over the stars which survive until the
present time (see also Thomas, Greggio \& Bender \cite{thom99}).

The observable in elliptical galaxies is the velocity dispersion
instead of the mass, so in order to properly compare our results with
observations we need to assume a correlation between the stellar mass
and the velocity dispersion of galaxies (Faber-Jackson relation).  We
assume:

\begin{equation}
\sigma = 0.86 M_{tot}^{0.22},
\end{equation}
\noindent
(Burstein et al. \cite{burs97}), where $\sigma$ is the velocity
dispersion in km/s.  The resulting relation between $\sigma$ and
$\Delta t$ can be seen from Fig. \ref{deltat} where we have indicated
in the upper scale the $\sigma$ corresponding to each SFR.

\begin{figure*}
\centering
\epsfig{file=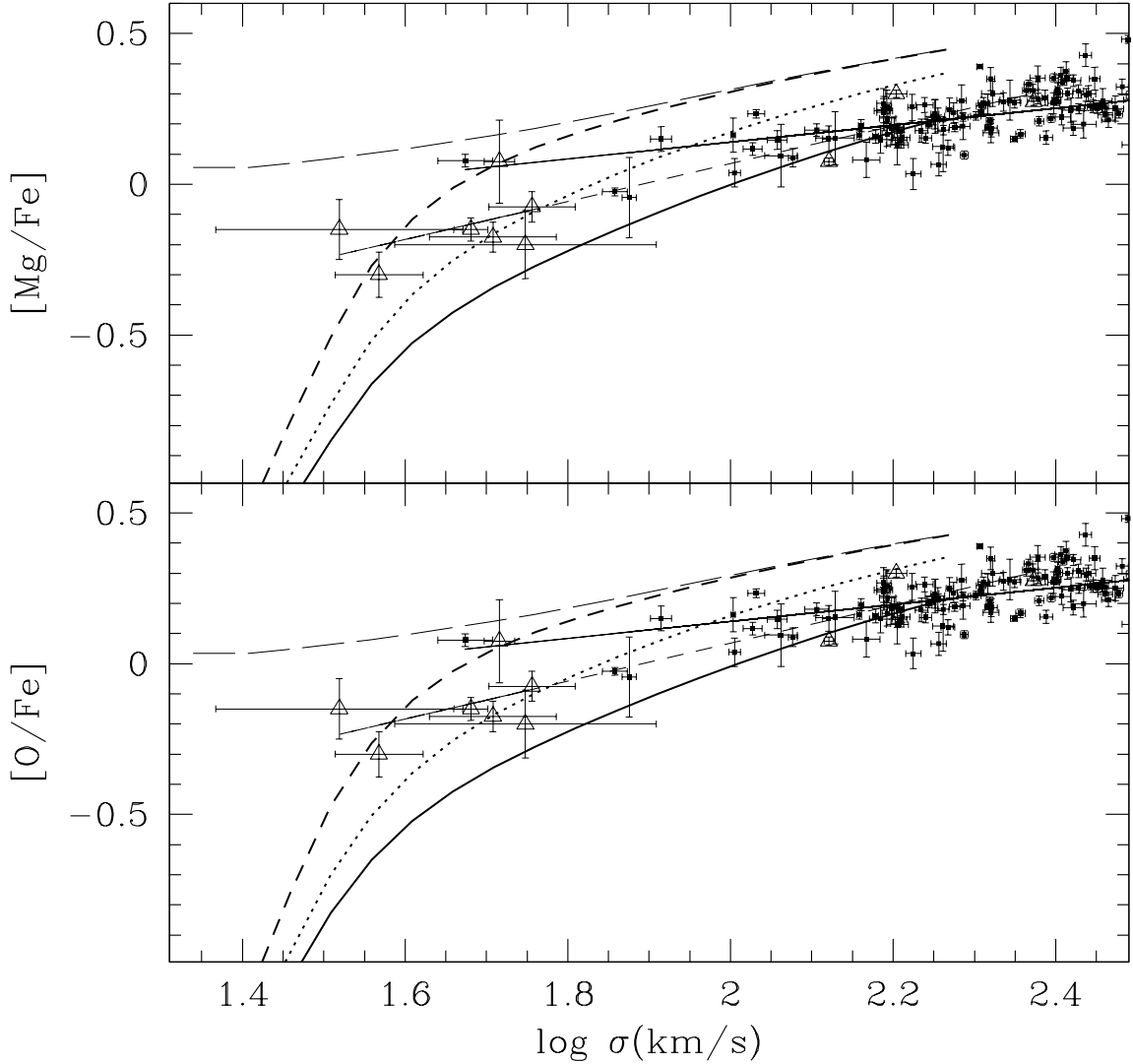, height=17cm,width=17cm}
\caption[]{\label{alphafe_gamma0.3} Mass-weighted [Mg/Fe] (upper 
panel) and [O/Fe] (lower panel) vs. $\sigma$ for models with a
constant SFR over a period of time $\Delta t$ which depends on the
stellar mass of the galaxy (see text) assuming $\gamma = 0.3$.  Heavy
solid line: model BETA235; heavy dotted line: model BETA200; heavy
short-dashed line: model BETA100; long-dashed line: model with a
constant (i.e. not SFR-dependent) canonical IMF.  The filled squares
are the observational values and relative error-bars as reported by
THOM05 and references therein and the light solid line is the
least-square fit of this data.  The open triangles are [$\alpha$/Fe]
ratios reported by Sansom \& Northeast (\cite{sn08}) and the light
short-dashed line is the corresponding least-square fit.}
\end{figure*}

In Fig. \ref{alphafe_gamma0.3} we show our results for $\gamma = 0.3$
comparing our models with observations taken from THOM05 and
references therein (filled squares).  We can first notice that, as
expected, the model BETA100 (heavy dashed lines), giving rise to
flatter IGIMFs (see Fig. \ref{igimf_ext}), produces larger
[$\alpha$/Fe] ratios.  In fact, flatter IGIMFs result in a larger
fraction of massive stars and, therefore, to a larger production of
$\alpha$-elements.  We can also appreciate that the models reproduce
quite well the [$\alpha$/Fe] (both [O/Fe] and [Mg/Fe]) ratios in
elliptical galaxies, at least for the models BETA200 and BETA235.  To
appreciate the effect of the IGIMF approach, we plot also (long-dashed
line) a model with the fixed canonical IMF which, as mentioned in
Sect. \ref{sec:igimf} and reminded in Sect. \ref{subs:SNIa}, has the
form $\xi(m) = k m^{-\alpha}$, with $\alpha = 1.3$ for 0.08 M$_\odot
<$ $m$ $<$ 0.5 M$_\odot$ and $\alpha = 2.35$ above 0.5 M$_\odot$.  The
curves obtained with the IGIMF tend to flatten out at large $\sigma$,
whereas the curve obtained with the constant IMF shows a constant
slope.  This demonstrates once more that the adoption of the IGIMF is
particularly remarkable in the low-mass (and low-$\sigma$) galaxies.
The curve with a constant IMF approaches asymptotically the model
BETA100 since this model at large SFRs produces the flattest IMFs (see
Fig. \ref{igimf_ext}).  Besides a small shift of a few tenths of dex
(which can be fixed increasing the parameter $A$ in Eq. \ref{eq:r1a}),
the curve with a constant IMF reproduces well the trend of
[$\alpha$/Fe] vs. $\sigma$ of the THOM05 sample, demonstrating that
the downsizing (or inverse-wind) models (Matteucci \cite{matt94};
Pipino \& Matteucci \cite{pm04}) are also perfectly capable of
explaining this trend in large elliptical galaxies.  However, evidence
is mounting that [$\alpha$/Fe] ratios in early-type dwarf galaxies are
solar or sub-solar.  For instance, van Zee, Barton \& Skillman
(\cite{vbs04}) showed that [$\alpha$/Fe] ratios (derived from Lick
indices) of a sample of Virgo dwarf irregular galaxies range between
-0.3 and solar.  Also in the cluster Abell 496 the smallest galaxies
show [Mg/Fe] to be solar or sub-solar (Chilingarian et
al. \cite{chil08}).  To show that we have also plotted in
Fig. \ref{alphafe_gamma0.3} (open triangles) the data of a sample of
low-mass early-type galaxies by Sansom \& Northeast (\cite{sn08}).
These data confirm that the [$\alpha$/Fe] vs. $\sigma$ relation is
probably steeper in the low-mass regime and that our IGIMF results can
naturally predict this behavior.  However, in order to properly test
our results in the low-mass regime more data are needed.

It is worth pointing out that in this figure (and in the following
ones) we have considered only model galaxies for which the SFR is
smaller than 100 M$_\odot$ yr$^{-1}$.  This is the reason why the data
points reach larger $\sigma$ than the results of our model.  In
extreme starbursts the IMF might become top-heavy as evident by the
mass-to-light ratios in ultra-compact dwarf galaxies which are
ultra-massive ``star clusters'' that form when the SFR is very high
(Dabringhausen, Kroupa \& Baumgardt \cite{dkb09}) and this will need
to be incorporated in the IGIMF calculations (work in preparation).

\begin{figure}
\centering
\epsfig{file=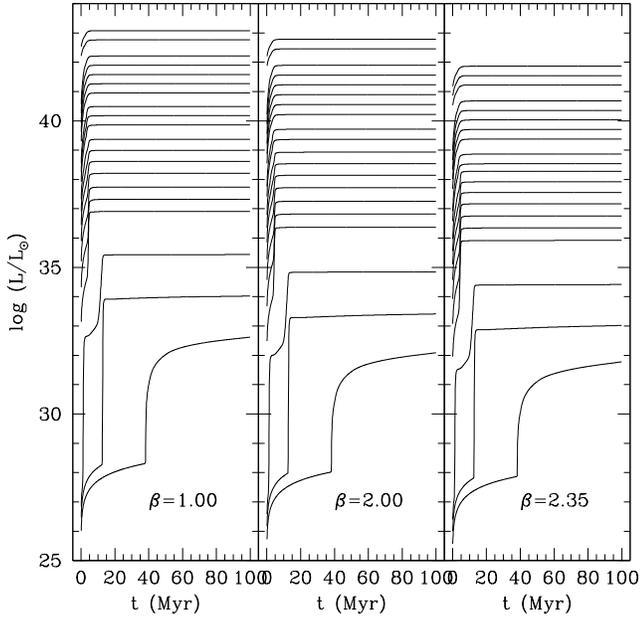, height=9cm,width=9cm}
\caption[]{\label{lum} Stellar luminosities (in L$_\odot$) as a function 
of time (in Myr) for models with constant SFR and different values of
$\beta$: model BETA100 (left panel), BETA200 (central panel), BETA235
(right panel).  In each panel, the lowermost curve has SFR = 10$^{-4}$
M$_\odot$ yr$^{-1}$, the uppermost has SFR = 10$^{2}$ M$_\odot$
yr$^{-1}$ and the other SFRs are equally spaced in logarithm.  These
luminosities have been obtained with the Starburst99 package
(Leitherer et al. \cite{leit99}; V{\'a}zquez \& Leitherer
\cite{vl05})}
\end{figure}

\begin{figure}
\centering
\epsfig{file=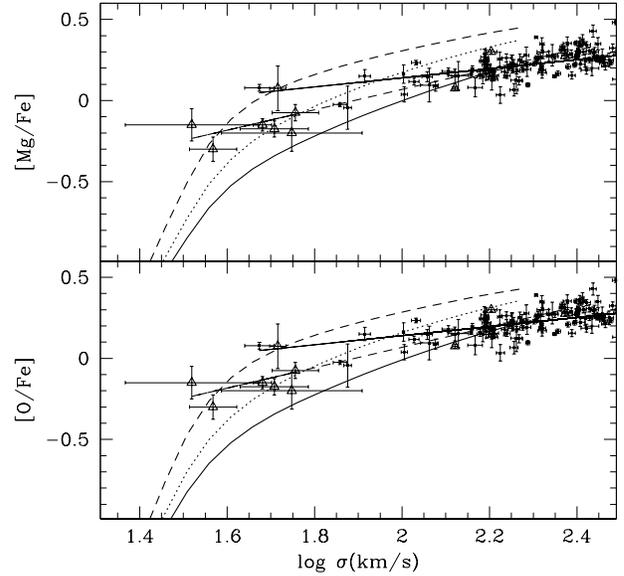, height=9cm,width=9cm}
\caption[]{\label{alphafe_lumw} As in Fig. \ref{alphafe_gamma0.3} 
but with luminosity-weighted [$\alpha$/Fe] ratios, calculated by 
means of eq. \ref{eq:alphafe_lw}.}
\end{figure}

In general, in local early-type galaxies the stellar abundances are
measured by means of various absorption-line Lick indices, such as
Mg$\, b$ and $<Fe> = 0.5 (Fe52720 + Fe5335)$ (THOM05). To properly
compare predictions to observational abundance data obtained for local
ellipticals, in general one should derive the luminosity-weighted
average abundances.  The real abundances averaged by mass are larger
than the luminosity-averaged ones, owing to the fact that, at constant
age, metal-poor stars are brighter (Greggio \cite{greg97}).  To
calculate the luminosities we have made use of the Starburst99 package
(Leitherer et al. \cite{leit99}; V{\'a}zquez \& Leitherer
\cite{vl05}), producing $L (t)$ for each value of SFR and $\beta$.
The results are shown in Fig. \ref{lum} for the first 100 Myr (the
luminosities remain almost constant after 100 Myr).  As expected,
since model BETA100 is characterized by the flattest IGIMFs, it
produces also the largest luminosities.  We have then calculated
luminosity-weighted mass ratios by using the formula:
\begin{equation}
[\alpha/Fe] = \log_{10} \frac{\int_{0}^{\Delta t} {\alpha \over Fe} (t) 
L (t) dt}{\int_{0}^{\Delta t} L (t) dt} - \log_{10}{\alpha_\odot
\over Fe_\odot}.
\label{eq:alphafe_lw}
\end{equation}
\noindent
The results are shown in Fig. \ref{alphafe_lumw}.  As we can see, the
results differ very little (by a few hundredths of dex at most)
compared to the mass-averaged abundance ratios.  We have checked these
results also using the spectro-photometric code of Jimenez et
al. (\cite{jime98}) (see also Calura \& Matteucci \cite{cm03}) but the
results do not differ appreciably compared with the ones obtained with
the Starburst99 package.  Indeed, it has been already shown in the
literature (but for constant IMFs) that the discrepancy for the
[Mg/Fe] ratio computed by averaging by mass and by luminosity is very
small, with typical values of 0.01 dex (Matteucci, Ponzone
\& Gibson \cite{mpg98}; Thomas et al. \cite{thom99}).  We have 
confirmed this finding also in the case of the IGIMF.

\begin{figure}
\centering
\epsfig{file=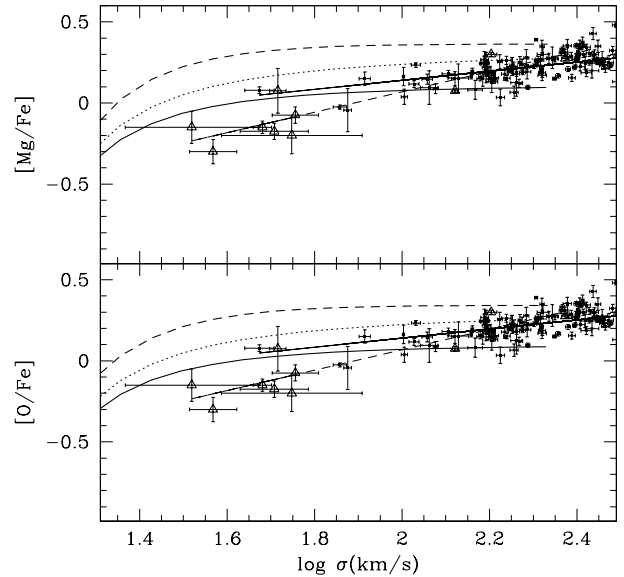, height=9cm,width=9cm}
\caption[]{\label{alphafe_dtc} As in Fig. \ref{alphafe_gamma0.3} 
but with $\Delta t$ = 1 Gyr for each stellar mass.}
\end{figure}

To check how much our results depend on the assumption of a variable
$\Delta t$ with stellar mass, we plot in Fig. \ref{alphafe_dtc} the
[$\alpha$/Fe] obtained assuming a constant value of $\Delta t$ = 1
Gyr.  The agreement with the observations is still quite good; in
particular the models maintain an increasing trend of [$\alpha$/Fe]
with $\sigma$.  However, the curves tend to flatten out too much at
larger $\sigma$, at variance with the trend shown by the observations.
This is due to the fact that, as pointed out in Sect. \ref{sec:igimf},
the various IGIMFs for rates of star formation larger than 1 M$_\odot$
yr$^{-1}$ do not show very large differences.  Therefore, the
assumption of a star formation duration inversely proportional to the
stellar mass of the galaxy (or in other words the downsizing) is a key
ingredient to understand the chemical properties of large elliptical
galaxies.

\begin{figure}
\centering
\epsfig{file=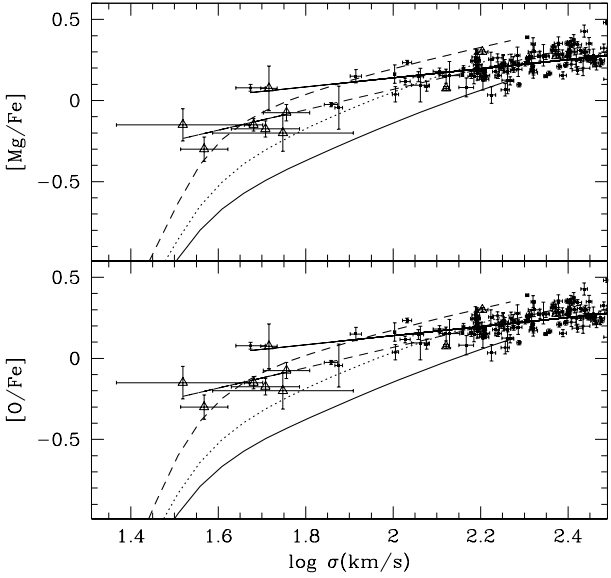, height=9cm,width=9cm}
\caption[]{\label{alphafe_gamma2.0} As in Fig. \ref{alphafe_gamma0.3} 
but with $\gamma$ = 2.0.}
\end{figure}

\begin{figure}
\centering
\epsfig{file=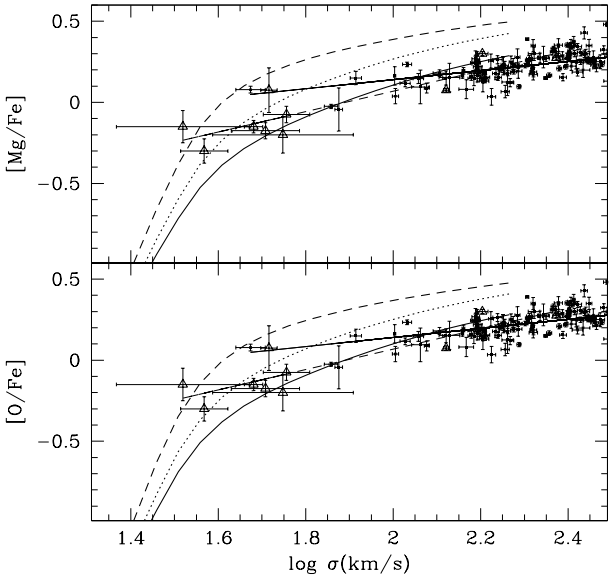, height=9cm,width=9cm}
\caption[]{\label{alphafe_gamma-0.3} As in Fig. \ref{alphafe_gamma0.3} 
but with $\gamma$ = -0.3.}
\end{figure}

To appreciate the dependence on the distribution function of mass
ratios in binary stars (the parameter $\gamma$ introduced in
Sect. \ref{subs:SNIa}) we plot in Figs. \ref{alphafe_gamma2.0} and
\ref{alphafe_gamma-0.3} the results of models with $\gamma$ = 2.0 and
$\gamma$ = -0.3, respectively.  The curves obtained with $\gamma$ =
2.0 tend to be slightly steeper than the ones shown in
Fig. \ref{alphafe_gamma0.3} (and slightly steeper than the
observations) but the agreement remains still good, in particular for
the models BETA100 and BETA200.  If we assume $\gamma$ = -0.3 an
excellent match with the observations is instead provided by the model
BETA235.  Models BETA100 and BETA200 show the same slope of the
observational data but shifted by a few tenths of dex.  A slight
increase of the parameter $A$ in eq. \ref{eq:r1a} would make these
models perfectly compatible with the observations.  

It is particularly remarkable that the trend of [$\alpha$/Fe] ratios
vs. $\sigma$ (namely an increase of [$\alpha$/Fe] with $\sigma$) is
naturally reproduced using the IGIMF approach, without any further
assumption or fine-tuning of parameters.  This is for instance at
variance with what hierarchical clustering models of structure
formation would tend to produce, since in this case larger elliptical
galaxies are formed later, out of building blocks in which the
[$\alpha$/Fe] ratio has already dropped (e.g.  Thomas et
al. \cite{thom02}).  It is worth mentioning that De Lucia et
al. (\cite{delu06}), by means of a semi-analytical model adopting the
concordance $\Lambda$ CDM cosmology, suggested that more massive
ellipticals should have shorter star formation timescales, but lower
assembly (by dry mergers) redshift than less luminous systems.  This
is one of the first works based on the hierarchical paradigm for
galaxy formation producing downsizing in the star formation histories
of early-type galaxies through the inclusion of AGN feedback (see also
Bower et al. \cite{bower06}; Cattaneo et al. \cite{catta06}), although
they did not compute the [$\alpha$/Fe]--$\sigma$ relation for
ellipticals.  However, the lower assembly redshift for the most
massive system is still in contrast to what is concluded by Cimatti,
Daddi \& Renzini (\cite{cdr06}), who show that the downsizing trend
should be extended also to the mass assembly, in the sense that the
most massive ellipticals should have assembled before the less massive
ones. Very recently, Pipino et al. (\cite{pipi08}) showed that even in
semi-analytical models able to account for the downsizing, the
[$\alpha$/Fe] vs. $\sigma$ relation is not reproduced.

\begin{figure}
\centering
\epsfig{file=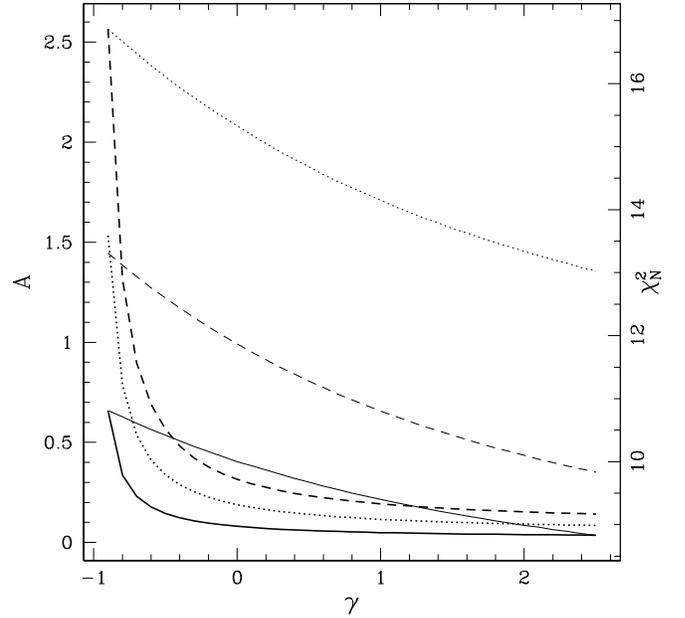, height=9cm,width=9cm}
\caption[]{\label{testA} Normalization constant $A$ to adopt in order to 
obtain the best fit with the observational data as a function of
$\gamma$ for model BETA100 (heavy dashed line), BETA200 (heavy dotted
line) and BETA235 (heavy solid line).  The light lines represent,
for each value of $\beta$, the normalized chi square of the represented 
model (scale on the right axis).}
\end{figure}

Although the agreement between our results and the observations
is good, none of the models presented so far fits perfectly the data
at low and high $\sigma$ simultaneously.  In order to work out an
overall best model, for each value of $\gamma$ and $\beta$ we have
checked, by means of a minimization of the normalized chi square,
which normalization constant $A$ fits better the data.  The results
are shown in Fig. \ref{testA}.  As we can see, model BETA235 seems to
be preferable and the best agreement between data and models is
obtained for the model BETA235 with $\gamma$ = 2.5 and $A$ = 0.036.
In general, the best fits are obtained with large values of $\gamma$,
although that requires low values of $A$.  A large value of $\gamma$,
favoring equal-mass binary systems, is consistent with the results of
Shara \& Hurley (\cite{sh02}), although observational surveys cited in
Sect. \ref{subs:SNIa} seem to indicate lower values of $\gamma$.

\begin{figure}
\centering
\epsfig{file=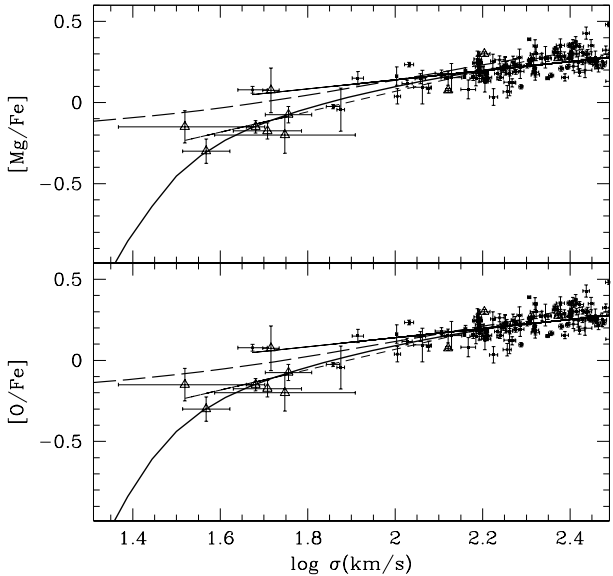, height=9cm,width=9cm}
\caption[]{\label{bestfit} Mass-weighted [Mg/Fe] (upper 
panel) and [O/Fe] (lower panel) vs. $\sigma$ for our best IGIMF model
(namely model BETA235 with $\gamma = 2.5$ and $A = 0.036$) with a
$\Delta t$--luminous mass relation described by eq. \ref{eq:newdtl}
(heavy solid lines).  Also shown (heavy long-dashed lines) a model
with fixed, canonical IMF, $\gamma = 2.5$ and $A = 0.13$.  Notations
and symbols as in Fig. \ref{alphafe_gamma0.3}}
\end{figure}

\begin{figure}
\centering
\epsfig{file=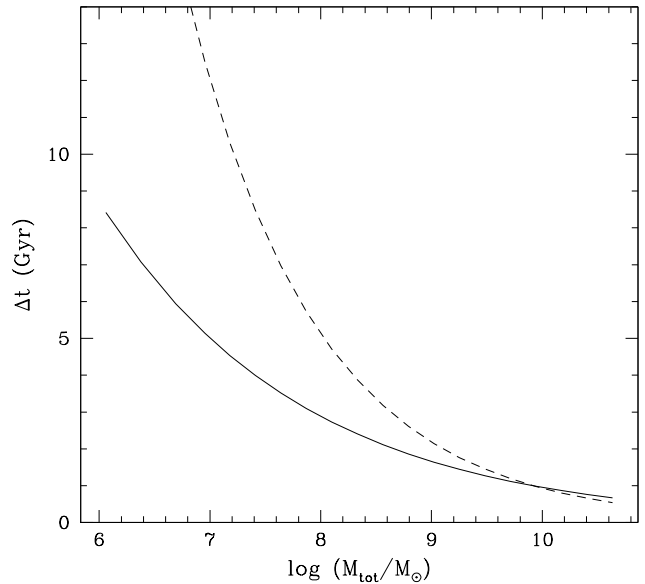, height=9cm,width=9cm}
\caption[]{\label{deltat_comp} $\Delta t$--luminous mass relation obtained 
with eq. \ref{eq:newdtl} (solid line) and derived by THOM05 (their
eq. 5; dashed line).}
\end{figure}

We should however not forget that the $\Delta t$--luminous mass
relation we have used in this work has been obtained by THOM05 {\it
assuming a constant IMF.}  We have therefore checked, starting from
our best model, namely a model with $\beta = 2.35$, $\gamma = 2.5$ and
$A = 0.036$, how this relation should change in order to best fit the
data.  It turns out that, within the IGIMF theory, the best $\Delta
t$--luminous mass relation is given by:

\begin{equation}
\log \Delta t = 2.38 - 0.24 \log M_{tot},
\label{eq:newdtl}
\end{equation}
\noindent
where $\Delta t$ is in Myr and $M_{tot}$ in M$_\odot$, and this
relation produces the results shown in Fig. \ref{bestfit}.  We have
also plotted in this figure a model with canonical stellar IMF (heavy
long-dashed line) in which we have increased the value of $A$ up to
0.13 in order to better reproduce the data.  We can notice again that
the canonical IMF can perfectly fit the [$\alpha$/Fe] ratios in large
ellipticals but it shows a constant slope and therefore it cannot
equally well reproduce the [$\alpha$/Fe] ratios in dwarf galaxies, at
variance with the IGIMF model.  The comparison between relation
\ref{eq:newdtl} and eq. 5 of THOM05 is displayed in
Fig. \ref{deltat_comp}.  We can notice here that the {\it downsizing
effect} (namely the shorter duration of the star formation in larger
galaxies) is milder, in the sense that the $\Delta t$ for large
galaxies is (slightly) larger than the timescale calculated by THOM05,
whereas at low $\sigma$ the star formation durations are significantly
lower than the ones predicted by THOM05.

\section{Discussion and conclusions}

In this paper we have studied, by means of analytical and
semi-analytical calculations, the evolution of [$\alpha$/Fe] ratios in
early-type galaxies and in particular their dependence on the luminous
mass (or equivalently on the velocity dispersion $\sigma$).  We have
applied the so-called integrated galactic initial mass function
(IGIMF; Kroupa \& Weidner \cite{kw03}; Weidner \& Kroupa \cite{wk05})
theory, namely we have assumed that the IMF depends on the star
formation rate (SFR) of the galaxy, in the sense that the larger the
SFR is, the flatter is the resulting slope of the IGIMF.  This kind of
behavior would naturally tend to form more massive stars (and
therefore more SNeII) in large galaxies, which are characterized by
more intense star formation episodes.  Therefore, it is expected that,
since $\alpha$-elements are mostly formed by SNeII, the most massive
galaxies are also the ones which attain the largest [$\alpha$/Fe]
ratios, in agreement with the observations.  One of the main aims of
this paper was to quantitatively check whether the chemical evolution
of galaxies within the IGIMF theory is able to accurately fit the
observed [$\alpha$/Fe] vs. $\sigma$ relation.

We have analytically calculated the SNII and SNIa rates with the IGIMF
assuming 3 possible slopes of the distribution function of embedded
clusters, $\xi_{\rm ecl} \propto M_{\rm ecl}^{-\beta}$, where $M_{\rm
ecl}$ is the stellar mass of the embedded star cluster; in particular
we have considered $\beta = 1.00$ (model BETA100); $\beta = 2.00$
(model BETA200); $\beta = 2.35$ (model BETA235).  We have seen that,
if we consider constant SFRs over the whole Hubble time, the final
SNIa and SNII rates agree quite well with the observations of spiral
galaxies (in particular the S0a/b ones).  The agreement with the
observed rates in irregular galaxies is not good, but a constant SFR
over the whole Hubble time is not likely in irregular galaxies, which
probably have experienced an increase of the SFR in the last Gyrs of
their evolution.

To calculate the [$\alpha$/Fe] ratios with the IGIMF we assumed
that early-type galaxies form stars at a constant rate over a period
of time $\Delta t$ which depends on the total luminous mass of the
considered galaxy.  This hypothesis is based on the work of THOM05
who, on the basis of observational grounds, showed the existence of a
downsizing pattern for elliptical galaxies, i.e. that the most massive
galaxies are the ones with the shortest $\Delta t$.  We have then
calculated the production of $\alpha$-elements and Fe by SNeII (in
particular we have calculated IGIMF-averaged SNII yields) and by SNeIa
and we have calculated mass-weighted and luminosity-weighted
[$\alpha$/Fe] ratios for each model galaxy, characterized by different
SFRs and $\beta$.

The resulting mass-averaged [$\alpha$/Fe] vs. $\sigma$ relations show
the same slope as the observations in massive galaxies as reported by
THOM05, irrespective of the value of $\beta$ and of the distribution
function of mass ratios in binaries $f(\mu) \propto \mu^{\gamma}$
(which affects the SNIa rates), although models with $\beta = 2.35$
and large values of $\gamma$ seem to be preferable.  Some models show
a shift (of a few tenths of dex) compared with the observations but
this can be fixed increasing (or decreasing) the fraction, $A$, of
binary systems giving rise to SNeIa, which is an almost unconstrained
parameter.  It is however remarkable that all the models we have
calculated show the same trend of the observations because if, as
commonly argued, large elliptical galaxies form out of mergers of
smaller sub-structures (hierarchical clustering), it would be natural
to expect that they are the ones with the lowest [$\alpha$/Fe] ratios
because they form later, out of building blocks where [$\alpha$/Fe]
has already dropped.

It is worth pointing out that the [$\alpha$/Fe] ratios do not depend
on the gas flows (infall and outflow) experienced by the galaxy
(Recchi et al. \cite{recc08}) therefore our results do not depend on
specific infall and outflow parameters, which make them particularly
robust.  However, these parameters affect the overall metallicity of
the galaxy, therefore they need to be taken into account in order to
check whether our models can correctly reproduce the mass-metallicity
relation.  As mentioned in the Introduction, K\"oppen et
al. (\cite{kwk07}) have already shown that the IGIMF theory is able to
reproduce the mass-metallicity relation found by Tremonti et
al. (\cite{t04}) in star-forming galaxies.  We are checking, by means
of detailed numerical models, that the IGIMF theory is able to
reproduce at the same time the mass-metallicity relation and the
[$\alpha$/Fe]-$\sigma$ relation in early-type galaxies.  This study
will be presented in a forthcoming paper.

We have also considered models in which the IMF does not vary with the
SFR and, because of the variations of $\Delta t$ with SFR, these
models are compatible with the observations of large elliptical
galaxies as well.  However, these models produce a [$\alpha$/Fe]
vs. $\sigma$ relation that can be described as a single-slope
power-law, whereas the IGIMF models bend down significantly at low
masses (and low $\sigma$).  This is because the IGIMF becomes
particularly steep in the galaxies with the mildest SFRs and this adds
to the downsizing effect (namely the decreasing duration of the SFR
with increasing mass).  From our study therefore, an important
conclusion is that a very reliable observable to test the validity of
the IGIMF theory is the observation of the [$\alpha$/Fe] ratios in
dwarf galaxies.  The available data on [$\alpha$/Fe] ratios in
low-mass early-type galaxies show indeed some steepening of the
[$\alpha$/Fe] vs. $\sigma$ relation, in agreement with the IGIMF
predictions.

We have also tested how much our results depend on the assumption of a
variable $\Delta t$ with stellar mass by computing models with $\Delta
t$ = 1 Gyr irrespective of the stellar mass.  The agreement between
models and observational data is still reasonably good but the curves
tend to flatten out too much at large stellar masses compared with the
observations (and with the IGIMF models).  This indicates that the
downsizing remains a fundamental ingredient to understand the chemical
properties of early-type galaxies.  However, if we check for which
$\Delta t$--luminous mass relation we obtain the best fit between data
and models, it turns out that the downsizing effect must be milder
than predicted by THOM05, in the sense that large galaxies form stars
for a slightly longer timescale than calculated by THOM05, whereas
low-mass galaxies have star formation durations significantly shorter.
Although the exact form of the best-fit $\Delta t$--luminous mass
relation is subject to a number of parameters (IGIMF parameters;
parameters regulating the SNIa rate etc.) and might change once larger
and more detailed abundance measurements are available, the result of
a milder downsizing effect compared to the findings of THOM05 is
robust.

Eventually, we have seen that luminosity-weighted [$\alpha$/Fe] ratios
agree very well with the mass-weighted ones (with relative differences
of a few hundredths of dex at most), in accordance with the results of
Matteucci et al. (\cite{mpg98}).  

We remind the reader that, with our analytical approach to chemical
evolution, we are making some important simplifying assumptions.  For
instance, our computation of the interstellar ${\alpha \over Fe} (t)$
given by eq. \ref{eq:alphafe_gas} does not take into account in detail
the lifetimes of massive stars. Furthermore, our present calculations
do not take into account the variation with time of the metallicity in
galaxies, which should also influence the stellar yields.  From the
various tests performed so far, and from the comparison of our results
with numerical results (Thomas et al. \cite{thom99}; Pipino \&
Matteucci \cite{pm04}), we have verified that these assumptions may
play only some minor role in determining the zero-point, but not the
slope of the predicted [$\alpha$/Fe] vs $\sigma$ relation.  All of
these simplifying assumptions will be relaxed in our forthcoming
paper, where we will present a numerical approach to the role of the
IGIMF in galactic chemical evolution.

The main results of our paper can be summarized as follows:

\begin{itemize}

\item Models in which the IGIMF theory is implemented naturally 
reproduce an increasing trend of [$\alpha$/Fe] with luminous mass (or
$\sigma$), as observed in early-type galaxies.

\item However, models with constant duration of the star formation 
produce a [$\alpha$/Fe] vs. $\sigma$ relation which flattens out too
much at large $\sigma$.  Only models in which the star formation
duration inversely correlates with the galactic luminous mass
(downsizing) can quantitatively reproduce the observations.

\item Models in which the IGIMF is implemented show (at variance with 
the constant IMF models) a steepening of the [$\alpha$/Fe]
vs. $\sigma$ relation for small galaxies, therefore the IGIMF theory
can be tested by observing the [$\alpha$/Fe] in dwarf galaxies.  The
observations available so far are in agreement with our predictions.

\item Luminosity-weighted abundance ratios differ from the mass-weighted 
ones by a few hundredths of dex at most.  This result, already known for 
constant IMF models, has been confirmed in the IGIMF framework.

\item In order to obtain the best fit between our results and 
the observed [$\alpha$/Fe] ratios in early-type galaxies, the downsizing
effect (namely the shorter duration of the star formation in larger
galaxies) has to be milder than previously thought.

\item The best results are obtained for a cluster mass function 
$\xi_{\rm ecl} \propto M_{\rm ecl}^{-2.35}$, indicating that the 
embedded cluster mass function should have a Salpeter slope.

\end{itemize}

\begin{acknowledgements}
      
We deeply thank Francesca Matteucci for help and support.  Discussions
with Antonio Pipino are also acknowledged.  S.R. acknowledges generous
financial support from the FWF through the Lise Meitner grant
M1079-N16.  F.C. and S.R. acknowledge financial support from PRIN2007
(Italian Ministry of Research) Prot. N. 2007JJC53X.  We thank the
referee, Daniel Thomas, for very useful comments.

\end{acknowledgements}

\end{document}